\documentclass[12pt]{article}
\usepackage{amsmath,amssymb,graphics,graphicx,multirow,verbatim,rotating,lscape}
\title{Hydrodynamics, horizons, holography and black hole entropy.}
\author{C. Sivaram\\
Indian Institute of Astrophysics\\
Bangalore 560034\\India\\
(sivaram@iiap.res.in)}
\date{}

\begin{document}

\maketitle

\section*{}
``Essay written for the Gravity Research Foundation 2011 Award for Essays on Gravitation''.\\
 \begin{center}
  Submitted on March 28, 2011.
 \end{center}

\section*{Summary}
The usual discussions about black hole dynamics involve analogies with laws of
thermodynamics especially in connection with black hole entropy and the associated
holographic principle. We explore complementary aspects involving hydrodynamics of the 
horizon geometry through the membrane paradigm. New conceptual connections complementing 
usual thermodynamic arguments suggest deep links between diverse topics like black hole 
decay, quantum circulation and viscosity. Intriguing connections between turbulence cascades,
quantum diffusion via quantum paths following Fokker- Planck equation and Hawking decay also
result from this combination of thermodynamic and hydrodynamic analogies to black hole dynamics.

\footnote{This essay was awarded an honourable mention by the Gravity Research Foundation
for the year 2011.}
\newpage
\section*{}
The holographic principle has its roots in black hole physics where there is
a unique convergence of general relativity, thermodynamics and quantum field theory
[1,2]. Classical general relativity implies that black hole mechanics is analogous to
the laws of thermodynamics, the black hole being assigned an entropy proportional to
its horizon area [3]. Hawking discovered that black hole radiates like a black body
with temperature $T$, which is inversely proportional to its mass $M$, thus fixing the black 
hole entropy, $S_{bh}$ as : [4]
\begin{equation}
 S_{bh}=\frac{A_H}{L^2_{Pl}}
\end{equation}
the horizon area, $\displaystyle{A_H=4\pi \left(\frac{2GM}{c^2}\right)^2}$ and $\displaystyle{L_{Pl}^2=\frac{\hbar G}{c^3}}$,
being the square of the Planck length. $G$, $c$ and $\hbar$ are respectively the Newtonian Gravitational constant,
velocity of light in vacuum and Planck's constant.

The holographic principle envisages the underlying microscopic degrees of freedom of gravity
in a region of bulk volume to be encoded on the boundary of the region. This makes the number
of microscopic degrees proportional to the area, not to the volume like in ordinary field or
particle theories without gravity. The conjecture is general and does not depend on the details
of the degrees of freedom or the emergent bulk gravity [5].

Quite apart from the well discussed thermodynamic analogies we also have the so  called membrane
paradigm in the classical general relativity [6], according to which any black hole has a fictitious
viscous liquid living on the horizon. Indeed the dynamics of the event horizon in the membrane paradigm
has been analyzed and shown to be described by the incompressible Navier-Stokes equation [7]. This 
implies a direct mapping between fluid variables of an incompressible flow and the geometrical 
variables pertaining to the horizon.

The fluid Navier-Stokes equations give an energy balance relating the change of kinetic energy to the rate of
decrease of viscous dissipation. In the corresponding geometric picture we can use the focusing equation relating
rate of change of kinetic energy to an increase in horizon area. In the geometrical picture, when the fluid is in the inertial range,
where viscous effects are small (high Reynolds number), the horizon area is 
nearly constant. We shall give direct examples and estimates in what follows.

Motions of Newtonian fluids are argued to be contained within a subclass of solutions of Einsteins equations
without sources [8]. A similar analysis can be done for relativistic fluids relating the hydrodynamic equations 
to horizon dynamics [9]. While thermodynamics describes static properties of a system in thermodynamic equilibrium,
hydrodynamics is an effective theory that describes long wavelength, small amplitude perturbations around thermal
equilibrium. Thus it requires knowledge of few parameters such as transport coefficients. One such coefficient is
shear viscosity, $\eta$, measuring momentum diffusion. It is roughly obtained from the energy momentum tensor, $T_{ij}\sim\eta\nabla_jT_{0i}$.
It is intuitive to think of weakly coupled systems as having low viscosities. In fact opposite is true,
particles in strongly coupled systems (with short mean free paths) have small viscosities.

It turns out as we shall see that both the viscosity and entropy density are related to the universal properties
of black hole horizons [10]. The analogy with membranes suggests that black hole horizons are characterized by a 
very low viscosity, i.e. the effective kinetic viscosity, $\nu$ is
\begin{equation}
 \nu\approx\frac{\hbar}{M}.
\end{equation}
The time scale, $t$ of decay of the system of scale length L, (vortex or eddy etc) is then given as:
\begin{equation}
 t\approx\frac{L^2}{\nu}.
\end{equation}
For black holes, $\displaystyle{L\approx \frac{GM}{c^2}}$ (horizon scale). Thus eqs (2) and (3) imply a decay time scale of
\begin{equation}
 t\approx\frac{G^2M^3}{\hbar c^4}.
\end{equation}
This is precisely the Hawking evaporation timescale. Thus the hydrodynamic analogy (horizon viscosity, etc)
suggests that the long timescale of black hole evaporation is due to the very low viscosity, larger black holes
having a smaller viscosity, the coefficient being inversely proportional to the mass $M$. This supplements the 
thermodynamic analogies. Indeed just as there is a universal upper bound on entropy, $S$ associated with black hole horizons
, i.e. $\displaystyle{S\leq\frac{A_H}{L_{Pl}^2}}$, for an enclosed region of surface area $A_H$, it has been established that the 
hydrodynamical behavior is better characterized by the ratio of shear viscosity to its entropy density, i.e. $\eta/S$,
(this ratio being a measure of viscosity) and this is characterized by a universal bound
\begin{equation}
 \frac{\eta}{S}=\frac{1}{4\pi}.
\end{equation}
This is a universal property of large $N$ strongly coupled finite temperature gauge theories with a gravity dual
and is independent of number of dimensions, or existence of chemical potentials. As remarked before, eq.(5) is
related to universal properties of black hole horizons [11,12].

[Briefly $\displaystyle{\eta=\frac{\sigma_{abs}}{16 \pi G} (\omega\rightarrow0)}$, i.e. zero frequency limit of absorption cross- section
of black hole, i.e. $\sigma_{abs}(\omega\rightarrow0)=A_H$, and as $\displaystyle{S=\frac{A_H}{4G}}$, we have: $\displaystyle{\frac{\eta}{S}=\frac{1}{4\pi}}$,
which is eq.(5)].

Eq. (5) implies a very low viscosity as compared to most substances in nature, being $380/4\pi$ for water and $9/4\pi$
for liquid helium. The above eqs. (1-5), imply that black holes decay into black holes of smaller mass, i.e. higher
viscosity, till it reaches the Planck mass $M_{Pl}$. Thus the hydrodynamic analogy suggests that black hole evaporation
is analogous to eddy dissipation cascade in turbulence, the smallest eddy at Planck scale being the most dissipative when
the `Reynolds number' becomes $\sim 1$, as we shall show below. As $\hbar/M$ is also the quantum of circulation, we consider
the total circulation of the black hole horizon (treated as a viscous fluid): i.e. we have the modified Feynman- Onsager
relation:
\begin{equation}
 \int_{R_H} \Omega R dR\simeq \frac{n\hbar}{M},
\end{equation}
where $R$ and $\Omega$ are the radial extent of the system and the angular velocity respectively.
Here for the black hole horizon, we have:$ \displaystyle{R=\frac{GM}{c^2}}$, $\displaystyle{\Omega=\frac{c^3}{GM}}$. etc.
Eq.(6), then gives the number of vortex lines, $n$ (each quantized vortex having circulation of
$\hbar/M$) as:
\begin{equation}
 n=\frac{GM^2}{\hbar c}
\end{equation}
on the horizon. This is analogous to rotation of a superfluid, where rotation goes into the 
vortices. This also implies a bound on the maximum angular momentum, $J_{max}$ of the black hole, i.e.
$\displaystyle{J_{max}<\frac{GM^2}{c}}$, as suggested by eq. (7). Eq. (5) also implies a vortex (surface)
density, independent of black hole mass. Eq. (7) also gives the entropy of the horizon.
The number of vortices decaying per second is $\displaystyle{\sim\frac{c^3}{GM}}$, so that the total decay time is 
(from eq. 7) again 
\begin{equation}
 t_d\sim \frac{G^2M^3}{\hbar c^4}
\end{equation}
which agrees with eq. (4)!
From the above parameters, the equivalent Reynolds number, $Re_{eff}$ (for the horizon fluid) characterizing
the decay is 
\begin{equation}
 Re_{eff}\approx\frac{GM^2}{\hbar c}.
\end{equation}
This is very large for large black holes. To draw the turbulence analogy the dissipation 
cascade would stop when $Re_{eff}\sim 1$, i.e., when the mass becomes of $\sim M_{Pl}$, the
most dissipative smallest eddy (low Reynolds number).
The above arguments and examples thus imply profound thermodynamical and hydrodynamical
analogies governing the quantum decay of black hole horizons. The classic boundary condition
associated with the horizon that fields may fall into the black hole and cannot emerge from it,
breaks time reversal symmetry and explain how Einstein equations can describe dissipative 
effects. (Baryon number conservation is also violated alongwith time reversal, in black holes,
this being equivalent to CP violation. This is discussed in [14]. As a turbulence analogy,
the Kolmogorov law, is not invariant under the reversal of velocity, $v\rightarrow- v$ and thus breaks time reversal symmetry.
Again emission of particles from the horizon can be viewed as a diffusive process [15], the
quantum paths being similar to diffusive trajectories of Brownian particles with a diffusion 
coefficient, $D$
\begin{equation}
 D=\frac{\langle x^2\rangle}{t}\sim\frac{\hbar}{M},
\end{equation}
where $\langle x^2\rangle$ is the root mean square displacement.
This gives a diffusion timescale, for a horizon scale $\displaystyle{L\sim \frac{GM}{c^2}}$ of again 
$t_{diff}\sim L^2/D$ which again is $\displaystyle{\frac{G^2M^3}{\hbar c^4}}$, agreeing 
with eq. (8). In quantum theory, the diffusion is connected to the finiteness of the 
Planck constant $\hbar$ (responsible for zero point motion) which in turn gives all the 
quantum properties of the horizon. A rigorous formalism involves using the analogy between the 
Fokker-Planck equation describing the spatial distribution $W$ of Brownian particles in a
potential $V(x)$ i.e.,$\displaystyle{\frac{\partial W}{\partial t}=D\frac{\partial^2 W}
{\partial x^2}-V(x)W}$, and the Schr\"{o}dinger equation 
$\displaystyle{\frac{\partial \psi}{\partial t}=\frac{-\hbar^2}{2M}\frac{\partial^2 \psi}
{\partial x^2}-V(x)\psi}$, describing the space time evolution of wave function, $\psi(x)$ of a quantum
particle in the potential $V(x)$. For $W(x,it)=\psi(x,t)$ (an example of Wick rotation) one gets $D=\hbar/M$.
It turns out that the Brownian paths dominate in the Feynman-Kac integral [16], expressed as
a product of oscillating potentials which interfere destructively for more irregular paths. 
This is a general result.

 Again as pointed out in refs. [17,18], the Casimir entropy of the 
horizon in terms of the summation over zero point modes also gives the usual entropy result.
The membrane analogy [15] further gives a black hole surface tension, $\displaystyle{T\sim \frac{c^4}{G} \frac{1}{GM/c^2}}$.
(This is related to the universal value of the superstring tension $\displaystyle{\frac{c^4}{G}}$ [19,20]). 

The pressure difference across a membrane or bubble of radius of curvature $R$ is:
$\displaystyle{P_{vac}^\prime=P_{vac}- \frac{2T}{R} }$ where $P_{vac}$ is the ambient vacuum pressure.
Inside the bubble $P_{vac}$ is modified as $\displaystyle{P_{vac}^\prime=P_{vac}\left(1- \frac{2GM}{Rc^2}\right)}$.
Any object of mass $m$ in the vacuum fluid displaces a volume, $V$ of $\displaystyle{
\frac{m}{\rho_{vac}}}$ or  $\displaystyle{\frac{m c^2}{P_{vac}}}$ (Archimedes effect!). So the pressure around the masses is
$\displaystyle{\simeq P_{vac}\frac{GM}{c^2 r}\frac{m c^2}{P_{vac}}\simeq\frac{GMm}{r}}$ and the force
(i.e. the pressure gradient) is just $\displaystyle{F=\frac{GMm}{r^2}}$, which is Newton's law.
Thus we have a hydrodynamic derivation of Newton's laws as force between bubbles immersed in a
sea of vacuum energy substratum. The modified expression for $P_{vac}$ also brings in relativistic effects.
Consequences for large scale structure and dark energy have been explored [21,22].

In short, we have explored conceptually new viewpoints to study features of black hole dynamics.
We find several intriguing correspondences between hydrodynamics, thermodynamics as well as
microscopic aspects like quantum diffusion and zero point energy underlying the evolution of
black hole horizons.

\newpage
\section*{References}
\begin{enumerate}
 \item L. Susskind: J. Math. Phys. $\mathbf{36}$, (1995), 6377.
\item G'. t Hooft: in A. Aly, J. Ellis et al. eds., World Scientific, (1993).
\item J. D. Bekenstein: PRD $\mathbf{7}$, (1973), 2333.
\item S. W. Hawking: Comm. Math. Phys. $\mathbf{43}$, (1975), 199.\\
C. Sivaram, K.P. Sinha: PRD $\mathbf{16}$, (1977), 75.
\item C. Sivaram, K. Arun: IJMP $\mathbf{D14}$,(2009),2167.
\item K. S. Thorne, R. H. Price, D. Mcdonald: Black holes: the membrane paradigm, Yale university press, (1986).
\item T. Damour: Proc. of 2nd Marcel Grossman meeting, ed. R. Ruffini, North Holland, (1982), 587.
\item I. Fouxon et al.: PRL $\mathbf{101}$, (2008), 261602.
\item S. Bhattacharya et al.: JHEP $\mathbf{0908}$, (2009), 059.
\item I. Fouxon, Y. Oz: arxiv:0909.3574(hep-th)
\item O. Aharony et al.: Physics Reports $\mathbf{323}$, (2000), 183.
\item I. R. Klebanov et al.: Physics Today $\mathbf{62}$, 2009, 28.
\item G. Falkovich et al.: RMP $\mathbf{73}$, (2001), 213.
\item C. Sivaram: Astr. Sp. Sci. $\mathbf{82}$, (1982), 485.
\item C. Sivaram: Lectures given in NATO Advanced Course on BH Physics, (1991), NATO-ASI Series, Kluwer, p225.
\item D. Sormette: Eur. J. Phys. $\mathbf{11}$, (1990), 334.
\item M. Rezven: J. Phys. A $\mathbf{30}$, (1997), 7783.
\item C. Sivaram: Asian J. Phys. $\mathbf{13}$, (2004), 293.
\item C. Sivaram: Astr. Sp. Sci. $\mathbf{167}$, (1990), 335.
\item C. Sivaram: IJTP, $\mathbf{33}$, (1994), 2407.
\item C. Sivaram: Astr. Sp. Sci. $\mathbf{219}$, (1994), 135.
\item C. Sivaram: Work in progress, (2011)
\end{enumerate}

\end{document}